\documentclass[prb,showpacs,superscriptaddress, twocolumn, preprintnumbers,amssymb]{revtex4}
\usepackage{graphicx}
\usepackage{dcolumn}
\usepackage{bm}
\usepackage{wasysym}
\usepackage{psfrag}
\usepackage{subfig}
\input epsf
\newcommand{\figurewidth}{84mm}
\begin{document}

\title{Comparison of two structures for transition-metal-based half Heusler alloys exhibiting fully compensated half metallicity}
\author{M. Shaughnessy}
\affiliation{Department of Physics, University of California, Davis, CA 95616-8677}
\affiliation{Lawrence Livermore National Laboratory, Livermore, CA 94551}
\author{C.Y. Fong}
\affiliation{Department of Physics, University of California, Davis, CA 95616-8677}
\author{L.H. Yang}
\affiliation{Lawrence Livermore National Laboratory, Livermore, CA 94551}
\author{C. Felser}
\affiliation{Institut fŸr Anorganische Chemie und Analytische Chemie, Johannes Gutenberg-UniversitŠt Mainz, 55099 Mainz, Germany}
\date{\today}
\begin{abstract}

We search for new fully compensated half metals, in which only one electronic spin channel is conducting and there exists no net magnetic moment. 
We focus on half Heusler alloys and we examine the physical consequence of different crystal structures found in the literature for 
these compounds, XMnZ, with a transition metal element, such as Cr, Mn, and Fe for X and a nonmetallic element, such as P, Sb and Si for Z. 
The structures differ in the placement of voids in the L2$_1$ structure of the full Heulser alloy. One structure has the void at 
(1/4, 3/4, 1/4)a and the other places the void at (0.0, 0.0, 1/2)a. The first structure is expected to have greater d-p hybridization between 
Mn and the Z atom. The other exhibits strong d-d hybridization between the nearest neighboring transition metal elements. Five XMnZ compounds are considered along with the previously studied CrMnSb in the second structure, which serves as a reference. Besides the CrMnSb, only one other alloy, MnMnSi, shows fully compensated half metallic properties in 
both structures. Both these alloys obey the Slater-Pauling electron counting rule for half Hesuler alloys.  The differences between
       CrMnSb and MnMnSi in the two structures are discussed based on
       their atomic properties.
In the search for fully compensated half metals in transition-
metal-based half Heusler alloys, we suggest using the counting rule as a guide.

\end{abstract}
\maketitle

\section{Introduction}\label{sec:AFM_intro}
             Impressive progress has been made in exploring so-called spintronic materials for fabricating efficient and non-volatile devices, 
such as spin valves\cite{Dieny}, spin current switches\cite{Bussmann}, and prototype magnetic random access memory (MRAM)\cite{Savtchenko}. 
In contrast with conventional transistors, these devices utilize the spin or spin and charge of an electron as the operational paradigm. Most of them 
are composed of layers of ferromagnetic transition metal (TM) elements and have imbalanced spin polarization at the Fermi energy, E$_F$\cite{Katine}. 
To be more effective, high spin polarization is particularly desirable. NiMnSb, a half Heusler alloy and a ferromagnet, has been predicted\cite{deGroot} 
to exhibit so-called half metallic (HM) properties Ð one of the spin channels shows metallic behavior while the oppositely oriented spin channel 
displays the characteristics of a semiconductor. The E$_F$ falls within the gap of the semiconducting channel. Thus, the spin polarization 
at E$_F$, P, is 100\%. Due to growth difficulties and various dynamic processes, such as surface effects and spin flip transitions\cite{Hordequin}, 
there has not yet been realized a device fabricated using a ferromagnetic half metal.

	Van Leuken et al.\cite{van Leuken} reported in 1995 a bulk TM-based (aside from the Mn atom) so-called fully compensated half-metal, CrMnSb, 
in the C1$_b$ structure of a half Heusler alloy, the same structure as de Groot's NiMnSb, and pointed out the absence of any net magnetic moment 
while still having fully spin polarized carriers. For device applications no net magnetization will be beneficial because there is no or low stray 
magnetic field to cause domain formations in nearby materials\cite{Nunez}. There has also been a suggestion that an interesting superconducting state 
should arise in a fully compensated half metal\cite{Rudd}. More recently, it was shown by a model Hamiltonian that small perpendicular current 
is needed to change the order parameter of layered structures made of antiferromagnetic materials\cite{Nunez}. Therefore, it is more appealing to 
use fully compensated half metals for spintronic applications. Most searches for fully compensated half metals have focused on theoretical 
studies of oxides, chalcopyrites, double pervoskites, and full Heusler alloys as summarized in \cite{Sasioglu}. They \cite{Sasioglu} confirmed 
the results of \cite{van Leuken} for CrMnSb and predicted that CrMnP and CrMnAs are also fully compensated half metals. In addition, they argue that 
the zero net magnetic moment is accounted for by the Slater-Pauling electron counting rule. In the original form, the Slater-Pauling rule is 
formulated by expressing every relevant quantity on a per atom basis. 

	For an ordered compound it is convenient to express the relevant quantities per formula or per unit-cell. For a half Heusler alloy, 
the Slater-Pauling rule gives a magnetic moment/formula-unit of:
\begin{equation}
M = (N_t  - 18)\mu_B 		
\end{equation}

where N$_t$ is the total number of electrons/formula unit, and $\mu_B$ is the Bohr magneton. The number 18 was suggested by K\"ubler\cite{Kubler} 
who took n$_{\downarrow}$, the number electrons having down spin orientation/atom, to be 3 in TM-based half Heusler alloys based on the results of 
calculated band structure for a total of 9 down spin bands occupied. These electrons are matched with spin up electrons to compensate their 
magnetic moments. For half Heusler alloys, the number of atoms/formula is 3. Then, (n$_{\downarrow}$ x 2) x 3 is 18. The three Cr-based half metals \cite{Sasioglu} 
all have 18 valence electrons/formula. So, M is zero.      
 
\begin{figure}
  \includegraphics[angle=0,width=3.6in,clip=true]{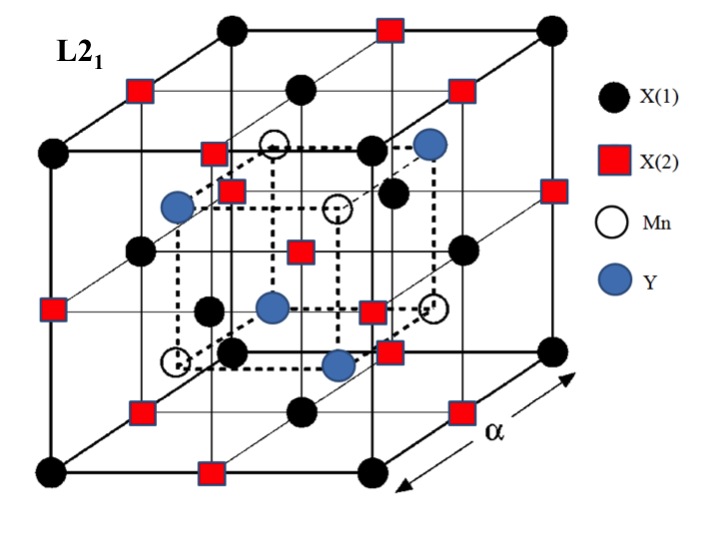}
  \caption{ The L2$_1$ structure. The C1$_b$ structure is defined when the set of either the X(2) squares (red) or the Y lightly shaded spheres (blue online) atoms is missing.
The former is S1 and the later is S2.}\label{fig:AFMstruct}
\end{figure}

	We found in the literature that the so-called C1$_b$ structure for half Heusler alloys can have three different arrangements of atoms inside the unit-cell. Let us start with the structure shown in Fig. \ref{fig:AFMstruct}, the L2$_1$ for a full Heusler alloy.  With either the X(2), square (in red), or Y, the lightly colored spheres (in blue), atoms removed, the new atomic arrangements inside the cubic cell form the C1$_b$ structure. Experimentally, the one determined by Mastronardi  et al \cite{Mastronardi} has the Y atoms missing. X(1), shown as a black circle, is occupied by a TM other than the Mn atom which is at (1/4,1/4,1/4)a, where a is the lattice constant of the face-centered cube. Either the pnictide or the group IV element occupies X(2) (square, red) site. Kandpal et al.\cite{Kandpal} used the structure similar to this one except the non-metallic atom at X(2) and the X(1) are switched. These two structures are close to the zinc-blende (ZB) structure for the III-V semiconductors having the non-metallic element as one of the nearest neighbors of the Mn atom and can in fact be transformed from one to the other by a reflection in the plane containing the Mn atoms. we call these two physically equivalent structures S1. The one used by Galanakis et al.\cite{Galanakis} and Van Lueken\cite{van Leuken} has the X(2) site as a void. The model was determined by Mancoff et al.\cite{Mancoff} when they grew superlattices of half Heusler alloys. We call this second structure S2. It is physically distinct from S1.
	
	According to Galanakis et al.\cite{Galanakis2}, the gap in the semiconducting channel of a half Heusler alloy in S2 is governed by the d-d hybridization. Based on earlier calculations of ferromagnetic half metals having the ZB structure, we expect that in S1 the d-p hybridization between the Mn and the nonmetal atom determines the bonding-antibonding gap.\cite{FongIonic} Therefore, we anticipate that the electronic and magnetic properties should reflect the difference of the atomic arrangements in the two structures. Furthermore, it will be interesting to examine if both structures strictly obey the Slater-Pauling electron counting rule and under which conditions a fully compensated half metal arises. 
	
	In this paper, we study six TM-based half Heusler alloys, namely, FeMnSi, FeMnP, CrMnSb, MnMnSi, CrMnSi, and VMnSb in the two structures. The previously studied CrMnSb in S2 serves as a reference.The first two compounds have more electrons, the last two have fewer, and MnMnSi has exactly the same number of electrons as the reference. We address the following issues: (A) Are all five new half Heusler alloys fully compensated half metals in both structures? For those that are fully compensated half metals in both structures, do they satisfy the Slater-Pauling rule? (B) Is there any evidence showing the two structures to have different physical properties? (C) For those alloys that are not fully compensated half metals, in particular those with odd number of electrons/primitive-unit cell, can the fully compensated half metallic state appear when using the conventional cell and imposing conditions forcing them to have zero moment? In section \ref{sec:AFM_method}, we comment on the two structures used in the calculations and briefly present the method of calculation. Results and discussion are given in Sec. \ref{sec:AFM_results}. Finally, we give a summary in Sec. \ref{sec:AFM_conclusion}.

           \section{Structures and Method} \label{sec:AFM_method}
           Because of issue (C) above, it is prudent to use the primitive unit cell to begin. Referencing to Fig. \ref{fig:AFMstruct}, the lattice vectors a, b, and c defining the primitive unit cell connect the X(1) atoms (black circles), at the face center of the cube, to the one located at the lower left corner, (0.0,0.0,0.0)a. This unit cell consists of three atoms. For S1, the non-metallic atom, the Mn, and the other TM are located at: (0.0,0.0,0.0)a, (1/4,1/4,1/4)a, and (0.0,0.0,1/2)a. The corresponding atoms in S2 are at (1/4,3/4,1/4)a, (1/4,1/4,1/4)a and (0.0,0.0,0.0)a. The conventional cubic cell shown in Fig. \ref{fig:AFMstruct} will also be used when we explore the possibility of expanding the unit cell for some of the five TM-based half Heusler alloys to yield a fully compensated half metal. It has a total of 12 atoms with four atoms for each species.
           
	We used the spin polarized version of the VASP algorithm.\cite{VASP} The PBE version of the generalized gradient approximation (GGA)\cite{Perdew} was used to treat the exchange-correlation between electrons. The ionic potentials of Cr, Fe, Mn, P, Si, Sb and V are constructed using the projector-augmented-wave (PAW) method with the GGA supplied by the VASP package. Plane waves were used as basis functions with cutoff energy between 800 - 1000 eV depending on the compounds. The Monkhorst and Pack\cite{Monkhorst} meshes of (15,15,15) were adopted. The convergence of the total energy and the magnetic moment of any sample are of the order of 1.0 meV and within 1.0 m$\mu_B$.

                 \section{Results and Discussion}\label{sec:AFM_results}
                 	We considered five half Heusler alloys, namely, CrMnSi, FeMnP, FeMnSi, MnMnSi and VMnSb as well as CrMnSb in S2; the last is one of the compounds considered in \cite{Sasioglu} and serves as a reference. Each compound is subjected to the respective atomic arrangements of the two structures. In the following, we address the issues raised in the Introduction. 
	
	The first issue is whether all the five compounds are fully compensated half metals in both structures. We examined cases with and without a scheme for forcing the total magnetic moment in the unit cell to be zero. We found that the fully compensated half metallicity can depend on the atomic arrangements inside the primitive unit-cell. In Tables \ref{tab:AFMHH1}
we show lattice constants at which the compounds to exhibit fully compensated half metallicity, total energies, and the values of the semiconducting gaps.

 \begin{table*}
\begin{center}
\begin{tabular}{cccc}
&& Structure 1 (2) &\\
Alloy & Lattice constant (\AA) &Total Energy (eV) & Semiconducting gap (eV) \\
\hline
CrMnSb & 5.87 (5.87) & -22.122 (-21.770) & 0.944 (0.690)  \\
CrMnSi & -- & -- & --  \\
FeMnP & -- & -- & --  \\
FeMnSi & -- & -- & -- \\
MnMnSi & 5.86 (5.86) & -22.780 (-22.780) & 0.740 (0.740)\\
VMnSb & -- & -- & -- \\
\hline
\hline
\end{tabular}
 \caption{Summary of half Heuslers in S1 and S2: the lattice constant at which the compound is fully compensated and half metallic, total energy, and the energy gap in the semiconducting channel. The data for S2 is in parentheses. Dashes are used for those alloys not exhibiting fully compensated half metallicity.}\label{tab:AFMHH1}
\end{center}
\end{table*}

	From the results shown in the table, MnMnSi is the only alloy with properties similar to CrMnSb, the reference Ð fully compensated half metallicity in both structures. Note that \cite{Sasioglu} considered CrMnSb only in S2.  

 Both CrMnSb and MnMnSi are fully compensated half metals over a range of lattice constants and fully compensated over a narrower range. Both MnMnSi and CrMnSb in either structure obey the Slater-Pauling electron counting rule. The fully compensated phases are stable as assured by their persistence both with and without a constraint on the total magnetic moment in the unit cell. We digress to mention that the VASP code allows for calculations to be done on a submanifold of wavefunctions subject to a constraint fixing the net magnetic moment in the unit cell to an arbitrary value. This constraining scheme is useful for quickly checking magnetic properties, but we have found that it can also give rise to unphysical states in which some states near E$_F$ have negative occupation. For all our calculations any imposed constraint on the total magnetic moment is removed at the end to allow the spin polarized wavefunctions to freely relax and avoid unphysical results. 
 
	There are three other features revealed in the table: (1) The total energy of CrMnSb in S1 is lower than in S2, while the total energies for MnMnSi are the same in both structures by symmetry. (2) The gap of CrMnSb in S1 is larger than MnMnSi under the same structure. On the other hand, the values are reversed in S2. (3) The gap of CrMnSb in S1 is larger than the one in S2 while MnMnSi shows the same gap.
	
	The first part of feature (1) can be attributed to the d-p hybridization occurring in S1 which opens a larger bonding-antibonding gap in the semiconducting (down spin) channel. The reason for feature (2) is that the electronegativity of Sb is larger than Si. In S2, because the Cr atom has one less electron than the Mn atom the strength of the d-d hybridization is reduced in CrMnSb. The first part of feature (3) provides more concrete evidence that the two structures exhibit different physical properties due to the d-p hybridization in S1 and d-d hybridization in S2. In S1, the Mn and the Sb forms the d-p hybridization. In S2, the Cr and the Sb could form the d-p hybridization. However, the Cr surrounded by eight atoms instead of four and the presence of the Mn serving as a second neighbor to the Sb atom, the d-p hybridization between the Cr and the Sb is weakened. In fact, the d-d hybridization between the Cr and the Mn atoms becomes important. For MnMnSi, the symmetry introduced by having the two identical TM atoms in the unit cell produces the same gap and also total energy in both structures. In S1, the hybridization is between the Mn at (1/4,1/4,1/4)a and the Si (0.0,0.0,0.0)a. The Mn at (0.0,0.0,1/2)a serves as second neighbor to the Si. In S2, the d-p hybridization is between the Mn at (0.0,1.0,0.0)a and the Si at (1/4,3/4,1/4)a. The Mn at (1/4,1/4,1/4)a is the second neighbor of the Si atom.

 
		
	In some cases antiferromagnetic states will be revealed only when using a larger unit cell \cite{Kittel} than is needed for non-magnetic or ferromagnetic states. In order to be sure we do not miss potential fully compensated half metals we also used the conventional cell for each of the alloys. FeMnSi is interesting. It has an odd number of electrons,19,  per primitive-unit-cell but shows fully compensated half metallicity in S1 when the magnetic moment is constrained to be zero. By analyzing in detail the states occupied in the semiconducting channel, we found that some of the occupied states are unphysical because of the constraint. The results of CrMnSi, FeMnP as well as VMnSb in both structures and FeMnSi in S2 are summarized in Table \ref{tab:AFMHH2}. Without the zero moment constraint, none are fully compensated half metals.

 \begin{table*}
\begin{center}
\begin{tabular}{cccc}
&& Structure 1 (2) &\\
Alloy & Lattice constant (\AA) &Total Energy (eV) & Semiconducting gap (eV) \\
\hline
CrMnSi & 5.7 (--) & --93.777 (--) & 0.674 (--)  \\
FeMnP & 6.55 (6.15) & -79.603 (-81.915) & 0.800 (0.811) \\
VMnSb & 5.95 (--) & -87.665 (--)  & 0.811 (--) \\
FeMnSi & 6.08 (6.45) & -83.271 (-77.466) & 0.937 (0.782) \\
\hline
\hline
\end{tabular}
 \caption{Summary of CrMnSi, FeMnP, VMnSb and FeMnSi in S1 and S2 for the conventional unit cell, subject to a zero magnetic moment constraint: the lattice constant at which the compound is fully compensated and half metallic, total energy, and the energy gap in the semiconducting channel. The data for S2 is in parentheses. For those models which are not half metallic dashes are given.}\label{tab:AFMHH2}
\end{center}
\end{table*}

	Two features in Table \ref{tab:AFMHH2} are worth noting: (i) Two half Heusler alloys, CrMnSi and VMnSb having one less electron/formula-unit than the number in CrMnSb and MnMnSi, 18, are not fully compensated half metalis in S2, and (ii) The alloys having one or more electron/formula-unit than CrMnSb can be fully compensated half metals in both structures when the conventional cell is used. Each of these half Heusler alloys should be ferromagnets according to the Slater-Pauling rule. For example, according to the rule M(VMnSb) = -4.0 $\mu_B$ (-1.0 $\mu_B$ in the primitive cell) and M(FeMnSi) = 4.0 $\mu_B$ (1.0 $\mu_B$ in the primitive cell). In fact, FeMnSi is numerically identified as a ferromagnet with M = 1.0 $\mu_B$ in S1 of the primitive cell. We may argue that if the Slater-Pauling rule gives a negative moment, the TM-based half metal cannot be fully compensated half metals in S2.  		
	
\subsection{Charge distributions}\label{sec:AFM_charge}

	   \begin{figure}
	  {
  \includegraphics[angle=0,width=\figurewidth,clip=true]{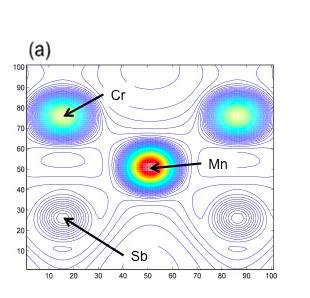}
    \includegraphics[angle=0,width=90mm,clip=true]{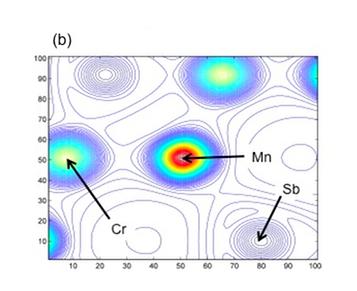}
    \caption{(a) CrMnSb in S1, (b) in S2; For S1, the section is formed by the [110] and [001] plane in the conventional cell.}
    \label{fig:AFM2a}}
    \end{figure}
    \begin{figure}{
      \includegraphics[angle=0,width=95mm,clip=true]{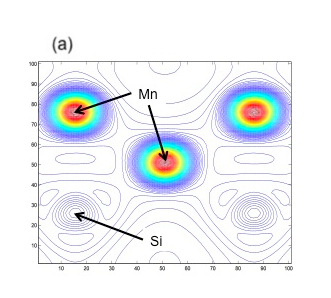}
    \includegraphics[angle=0,width=95mm,clip=true]{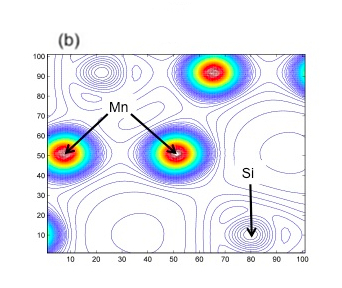}
  \caption{(a) MnMnSi in S1, (b) in S2. For S1, the section is formed by the [110] and [001] plane in the conventional cell.}\label{fig:AFM2b}}
\end{figure}  

		As stated, in S1 the d-p hybridization should be the dominant bonding feature of CrMnSb and MnMnSi while in S2 the d-d 
hybridization between the TMs form the bonds. In Figs. (\ref{fig:AFM2a}) and (\ref{fig:AFM2b}), we show the charge density plots of the two crystals in the sections containing 
the three atoms, two TMs and one non-metallic element. For S1, (Figs. \ref{fig:AFM2a}(a) and \ref{fig:AFM2b}(a)), the section can be easily labeled with axes along the [110] and [001] 
directions of the conventional cell.  
	The charge densities show small differences due to the differing half Heusler structures, for CrMnSb, and are equivalent, as expected, 
for MnMnSi. The value for the highest closed contour directly between the Mn and the Sb is approximately twice as large in 
 S2 as compared to S1. This demonstrates the strong d-p hybridization for S1.

\subsection{Density of states}\label{sec:AFM_DOS}
\begin{figure}
 {
    \includegraphics[angle=0,width=\figurewidth,clip=true]{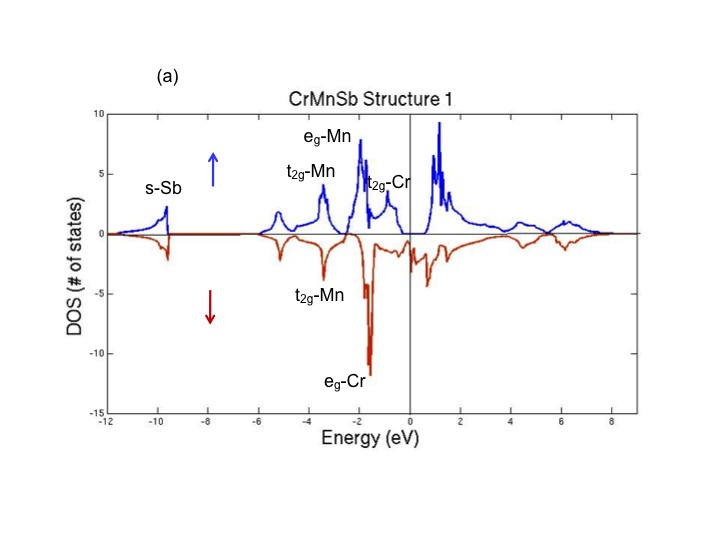}
    \includegraphics[angle=0,width=\figurewidth,clip=true]{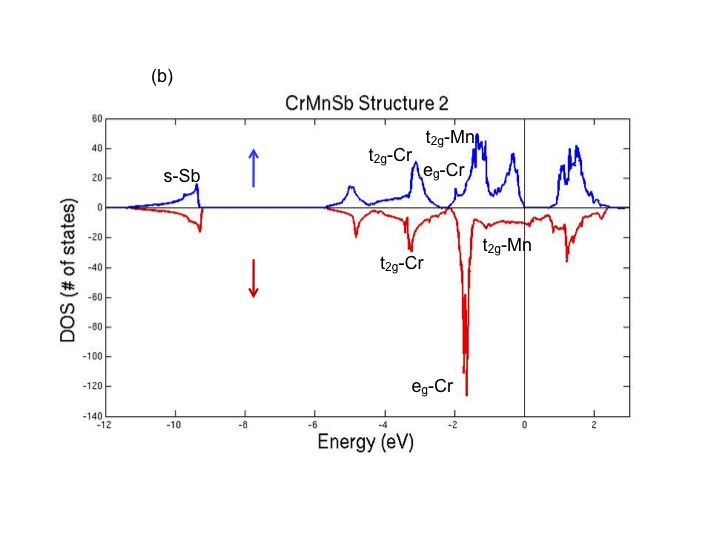}}
    \caption{(a) The density of states of CrMnSb in S1 and (b) in S2. }\label{fig:DOSAFM}
      \end{figure}

\begin{figure}
   {
  \includegraphics[angle=0,width=\figurewidth,clip=true]{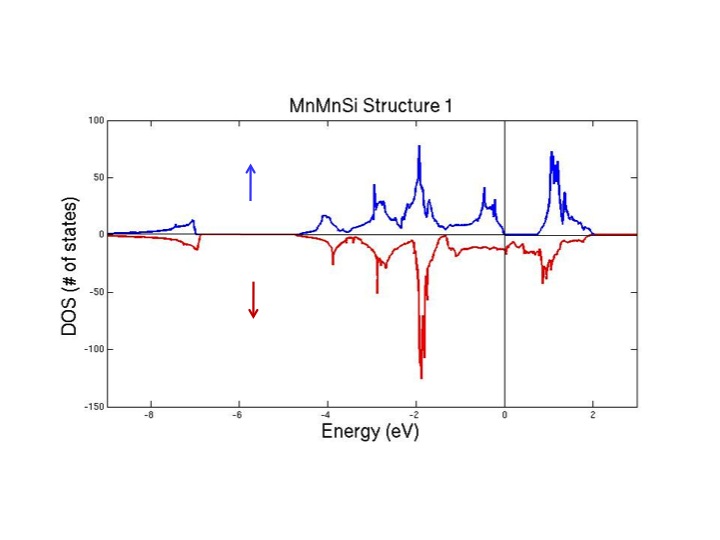}
    \includegraphics[angle=0,width=\figurewidth,clip=true]{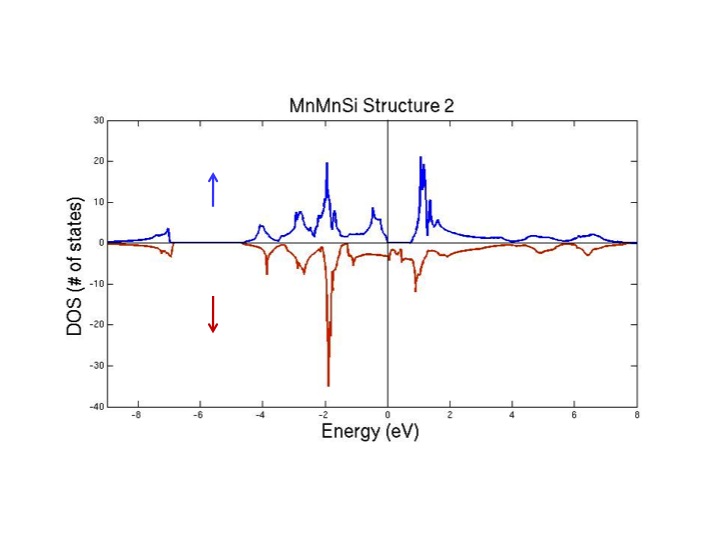}}
  \caption{ (a) The DOS of MnMnSi in S1 and (b) in S2. E$_F$ is set to zero.}\label{fig:DOSAFM2}
  \end{figure}

The density of states of CrMnSb in both structures are shown in Figs. \ref{fig:DOSAFM} and \ref{fig:DOSAFM2}. The number of peaks in both structures is the same for the two spin channels. The lowest energy peak is from the s-state of the Sb atom.

The dominant states of other structures in the occupied states are identified. The major effects of the structural difference are: (i) The e$_g$ state of the Cr atom. In S2, the bands associated with the e$_g$ states are narrower in the down spin channel. (ii) In the up spin channel, the t$_{2g}$ states of Mn and Cr are reversed. The lowered energy of these states of the Mn atom is due to the d-p hybridization to form the bonding states. 

	The density of states of MnMnSi are very similar to the ones shown in Fig. \ref{fig:DOSAFM}. The lowest states are now from the s-states of the Si atom. When Cr is replaced by Mn, the density of states for the MnMnSi is obtained. The missing electron of Si with respect to the Sb is reflected in the bandwidth with respect to E$_F$. The Sb-based compound has a larger width. In MnMnSi, they are located at about -8.0 eV. The extra d-electron of Mn has less effect on majority spin channel. But it causes the primary peak of the minor spin channel to shift down by 0.2 eV and to narrow the width for the states between -4.0 and -2.0eV as compared to CrMnSb.

           \section{Conclusion}\label{sec:AFM_conclusion}
                      The first publication to consider a half Heusler in a fully compensated electronic state\cite{van Leuken} and the recent model study\cite{Nunez} suggest there are advantages to using these zero moment half metals for spintronic applications. Half Heusler alloys are good candidates. We found that there exist two physically distinct structures for the half Heusler alloys in the literature. S1 is similar to the ZB structure and allows the d-p hybridization to be dominant. S1 will generally have lower energy due to the opening of the bonding - antibonding gap. S2 favors the d-d hybridization between the TM elements. Among the five studied half Heusler alloys, we only found one new compound, MnMnSi, exhibiting fully compensated half metallicity in both structures. 
                      
                      From an experimental prospective, this material may be difficult to grow, perhaps crystallizing in the Cu$_2$Sb structure or even something hexagonal. We note the Cu$_2$Sb structure is the native one for Fe$_2$As and Mn$_2$As\cite{Pearson}. Using a suitable substrate, such as Si or GaAs may allow for the epitaxial growth of a few layers of fully compensated MnMnSi in the half Heusler structure, since the spin compensation exists over a range of lattice constants. It may not be unexpected to find the spin compensation in the half Heusler MnMnSi because the half-filled d-shells of the Mn often give rise to antiferromagnetic order in confined volumes of crystalline solids. The spins of the d-states on individual Mn atoms are aligned in accordance with Hund's first rule, while the Pauli exclusion principle encourages anti-alignment of the local magnetic moments of neighboring atoms. This compound and the reference alloy, CrMnSb, do satisfy the Slater Ð Pauling electron counting rule. We suggest to apply this rule as a guide for successful search of TM based HM-AFM half Heusler alloys. The properties between S1 and S2 are compared. We also examine the effects of the valence of the nonmetal element and of an extra electron of the Mn in MnMnSi.

Work at UC Davis was supported in part by the National Science Foundation Grant No. ECCS-0725902. MS is supported by Lawrence Livermore National Laboratory, This work was performed under the auspices of the U.S. Department of Energy by Lawrence Livermore National Laboratory under Contract DE-AC52-07NA27344. LHY is supported by DOE under contract No. W-7405-ENG-48.

\end{document}